\documentclass[twocolumn]{revtex4}
\usepackage{amssymb}
\usepackage{amsfonts}
\usepackage{graphicx,amsmath}
\usepackage{setspace}

\begin{document}

\title{Non-mean-field screening by multivalent counterions}

\author{M. S. Loth\cite{contact} and B. I. Shklovskii}

\affiliation{Department of Physics, University of Minnesota,
Minneapolis, MN 55455}

\date{\today}

\begin{abstract}

Screening of a strongly charged macroion by its multivalent
counterions can not be described in the framework of a mean-field
Poisson-Boltzmann (PB) theory because multivalent counterions form a
strongly correlated liquid (SCL) on the surface of the macroion. It
was predicted that a distant counterion polarizes the SCL as if it
were a metallic surface and creates an electrostatic image. The
attractive potential energy of the image is the reason why the
charge density of counterions decreases faster with distance from
the charged surface than in PB theory. Using the Monte Carlo method
to find the equilibrium distribution of counterions around the
macroion, we confirm the existence of the image potential energy. It
is also shown that due to the negative screening length of the SCL,
$-2\xi$, the effective metallic surface is actually above the SCL by
$|\xi|$.

\end{abstract}

\maketitle

\section{Introduction}\label{sec:Introduction}

In this paper we deal with a problem in which one big and strongly
charged ion, called a macroion, is screened by much smaller but
still multivalent counterions, each with a large charge $Ze$ ($e$ is
the proton charge); for brevity, we call them $Z$-ions. A variety of
macroions are of importance in chemistry and biology, including
charged lipid membranes, colloids, DNA, and viruses. Multivalent
metal ions such as La$^{+3}$, dendrimers, and short polyelectrolytes
can play the role of the screening $Z$-ions.

\begin{figure}[h!]
\includegraphics*[scale=0.55, viewport=22 4 462 278]{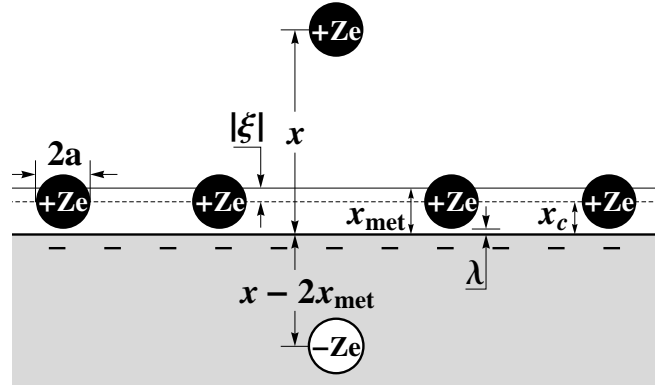}
\caption{A stray positive $Z$-ion (elevated black sphere), at a
distance $x$ from the surface (thick line) of a negatively charged
planar macroion (shaded region). Other $Z$-ions (black spheres),
condensed at the surface, are on average a distance $\lambda$ above
the macroion's surface. The dashed line indicates the average
distance, $x_{c}=a+\lambda$, from the macroion's surface to the
adsorbed $Z$-ions' centers. The stray $Z$-ion and its negative image
charge (white sphere), are equidistant from the effective metallic
surface, which is shown by the thin line at $x_{met}=x_{c}+| \xi
|$.} \label{fig:CorrHolePlane}
\end{figure}

To illustrate the fundamental aspects of screening we use the simple
geometry of a solid occupying the half space $x\leq 0$, whose
surface at $x=0$ has a large uniform surface charge density
$-\sigma$. The surface charge is screened by an aqueous solution of
positive, spherical $Z$-ions with radius $a$, which occupies the
rest of space $x > 0$ (see Fig. \ref{fig:CorrHolePlane}). Both the
macroion and the aqueous solution have dielectric constant $\epsilon
\simeq 80$. If all of the $Z$-ions were to condense on the
macroion's surface, their total charge per unit area would equal
$\sigma$. In such a neutral system, the concentration of $Z$-ions
$N(x)\rightarrow 0$ at $x\rightarrow \infty$. The main goal of this
paper is to discuss the behavior of $N(x)$. The solution of the
Poisson-Boltzmann (PB) equation for this problem has been known for
nearly a century~\cite{Gouy1910, Chap1913}. The Gouy-Chapman
solution is
\begin{equation}
N(x)=\frac{1}{2 \pi Z^{2} l_{B} }\frac{1}{(\lambda + x - a)^{2}},
\label{eqn:GC}
\end{equation}
where $\lambda = e/2 \pi \sigma  l_{B} Z $ is the Gouy-Chapman
length, and $l_B=e^{2}/(\epsilon k_{B} T)\simeq 0.71$ nm is the
Bjerrum length. We have modified the standard Gouy-Chapman formula,
taking into account the finite radius of the $Z$-ions, which can not
approach the surface closer than $x=a$.

It was shown~\cite{PS99, S99, Colloquium02, L2002, N_rev2005,
Mess_rev2008, MN2000, BAO2004, MN2001} that the Gouy-Chapman
solution fails when both $\sigma$ and $Z$ are large enough. The
reason it fails is that, in addition to $\lambda$, there is a second
length scale in the problem due to the discreteness of charge.
When the condensed $Z$-ions neutralize the charge of the plane, the
two-dimensional concentration of $Z$-ions is $n = \sigma / Ze$, and
the surface area per ion, the Wigner-Seitz cell, can be approximated
as a disc of radius $b$ such that $\pi b^2 = 1/n$. Thus, $b = (\pi n
)^{-1/2} = ( Z e / \pi \sigma )^{1/2}$ and $2b$ is approximately the
distance between Z-ions. We can construct the dimensionless ratio
\begin{equation}
\frac{b}{\lambda} = 2 \Gamma \ , \ \ \ \Gamma = \frac{Z^2 e^2
/\epsilon b}{ k_B T} \ . \label{eq:Gamma}
\end{equation}
Here $\Gamma$ is the dimensionless Coulomb coupling constant, or the
inverse dimensionless temperature measured in the units of a typical
interaction energy between $Z$-ions. For example, at $Z=3$ and DNA
like $\sigma=0.95 e$ nm$^{-2}$ used in this paper, we get $\Gamma =
6.4$, $\lambda \simeq 0.79$ nm and $b \simeq 1.0$ nm. Thus, the
Coulomb repulsion energy of the $Z$-ions dominates the thermal
energy. The result is a strongly correlated liquid, which has
short-range order similar to a Wigner crystal~\cite{PS99, S99,
Colloquium02, L2002, N_rev2005, Mess_rev2008, MN2000, BAO2004,
MN2001, ROUZ1996, GRON97, S1999_2} and is located, practically, at
the very surface of the macroion. This paper deals only with the
strong coupling case: $\Gamma \gg 1$. Another definition for a
Coulomb coupling parameter, $\Xi=2\Gamma^{2}$, was introduced in
Ref.~\cite{MN2000}, and of course, in the limit $\Gamma\gg1$,
$\Xi\gg1$ as well.

Mean-field treatments, along the lines of PB theory, fail at $\Gamma
\gg 1$, since when a $Z$-ion strays away from the plane to distances
$x-a\ll b$, the electric field of his neighbors has no significant
$\widehat{x}$ projection. In this range, the stray $Z$-ion is only
affected by the electric field of its Wigner-Seitz cell (a disc of
radius $b$). Therefore, at $x-a\ll b$, the surface charge of the
macroion is unscreened, and the electric field is
$2\pi\sigma/\epsilon$. Thus, for $0 < x-a \ll b$,
\begin{equation}
N(x)=\frac{\sigma}{Ze\lambda} \exp[-(x - a)/\lambda].
\label{eqn:close}
\end{equation}
(Here following Ref.~\cite{Colloquium02} we used an expression for
$N(a)$ that ignores the atomic structure of water, while
Refs.~\cite{PS99, S99} tried to take this structure into account).

Remarkably, the same length $\lambda$ characterizes both this
exponential decay and the Gouy-Chapman solution, Eq. (\ref{eqn:GC}).
It is clear that the dramatic difference between the exponential
decay of Eq. (\ref{eqn:close}) and the power law decay of Eq.
(\ref{eqn:GC}) is due to the effects of correlations. Eq.
(\ref{eqn:close}) was first obtained in Refs.~\cite{PS99,S99}. Then
it was re-derived in Refs.~\cite{MN2000,BAO2004} and confirmed by
Monte Carlo (MC) simulations in Ref.~\cite{MN2001}. Below we again
confirm Eq. (\ref{eqn:close}) at $0< x-a\ll b$ by MC simulations.
However, the focus of this paper is on the non-PB behavior of $N(x)$
at larger distances $x-a > b$, which has been predicted in
Refs.~\cite{PS99,S99} but to our knowledge has never been verified
analytically or numerically.

To bring this prediction to mind, let us focus on a single, stray
$Z$-ion located above the macroion's surface at $x>a+b$ (Fig.
\ref{fig:CorrHolePlane}). Refs.~\cite{PS99,S99} argue that the
negative charge of the correlation hole, $-Ze$, will spread to a
disc of size $\sim x$ as neighboring $Z$-ions move to occupy the
Wigner-Seitz cell the stray $Z$-ion left behind. This is similar to
what happens in a metallic surface under the influence of an
external charge. In fact, this metal-like polarization of the SCL on
the surface of the macroion can actually be described by an image
charge that appears in the body of the macroion. Because the centers
of the $Z$-ions which form the SCL are typically located at a
distance $x_c =a+ \lambda$ above the surface (see Fig.
\ref{fig:CorrHolePlane}) it is natural to think that the effective
metallic surface is at $x_{met} = x_c$ and therefore the image is
located at $-x + 2x_{met}$. The attractive interaction energy
between the stray $Z$-ion and its image is then~\cite{LL84}, for
$x-x_{met}\gtrsim b$,
\begin{equation}
U_{im}(x)=-\frac{Z^{2} e^{2}}{4 \epsilon (x-x_{met})}.
\label{eqn:PlaneImage}
\end{equation}
This attractive image interaction, of course, is a correlation
effect.

The goal of this paper is to verify, by a Monte Carlo (MC)
simulation and an analytical calculation, that a SCL on the
insulating surface of a macroion does behave as a metal, and a stray
$Z$-ion has potential energy $U_{im}(x)$. The plan of the paper is
as follows. In Sec. \ref{sec:MCSimulation} we describe our MC
procedure. In Sec. \ref{sec:MCResults}, we present our MC results
for the screening of a spherical macroion by $Z$-ions. To a first
approximation they confirm that a stray $Z$-ion at $x
>a+b$ has potential energy $U_{im}(x)$. This is the most important
result of our paper.

At a more detailed level, we see in Sec. \ref{sec:MCResults} that to
more accurately fit Eq. (\ref{eqn:PlaneImage}) to our MC data the
effective metallic surface must be lifted slightly above $x_{c}$. We
find that a shift of $0.21$ nm provides the best fit. This shift is
explained in Sec. \ref{sec:Location} where we analytically derive
Eq. (\ref{eqn:PlaneImage}), showing that there is indeed an
attractive interaction energy between the stray $Z$-ion and its
image. We further prove that the effective metallic surface should
be lifted slightly by $-\xi=|\xi|$, where $2\xi$ is the linear
screening length of the SCL. In other words, $x_{met}$, in Eq.
(\ref{eqn:PlaneImage}) should be replaced by $x_{met}=x_c -
\xi=x_{c}+|\xi|$. We show that, theoretically, $\xi = - 0.20$ nm, in
reasonable agreement with the MC simulation. The fact that a
Wigner-crystal-like SCL has a negative screening radius was
predicted theoretically~\cite{BLES1981} and confirmed experimentally
for a low-density two-dimensional electron gas in silicon MOSFETs
and GaAs heterojunctions~\cite{KRAV1990,Eis1992} (see also a recent
paper ~\cite{Efros08}).

In Sec. \ref{sec:addedsalt} we add a small concentration of
monovalent salt (for example, NaCl) to our system. We show that the
attractive image interaction persists in this system; however, the
attraction is weaker due to screening.

\section{Monte Carlo Simulation}\label{sec:MCSimulation}

Our setup is similar to the simulations found in
Refs.~\cite{MN2001,holm07,Diehl06,Diehl08,Perez05,Perez08}. Our
system is contained within a spherical cell with radius
$r_{max}=10.0$ nm. Centered within the cell is a spherical macroion
with charge $Q_{M} = -300 e$ and radius $R_{M}=5.0$ nm
($-\sigma=-0.95$ $e/$nm$^{2}$). The system is populated by $100$
$Z$-ions of charge $3e$ and radius $a=0.4 $ nm. The mobile particles
are initialized to random non-overlapping coordinates. The wall of
the spherical cell has a distance of closest approach of $0$ so that
a $Z$-ion may be placed with its center at the wall. Therefore all
$Z$-ions are found at a radial distance $r$ within the range
$R_M+a<r\leq r_{max}$. After initializing the system, the total
electrostatic energy of the system is calculated as
\begin{equation}
\mathcal {E}=\frac{e^{2}}{2 \epsilon} \sum_{i,j; i\neq
j}^{101}\frac{q_{i} q_{j}}{d_{ij}}, \label{eqn:coulomb}
\end{equation}
where particle $i$ has charge $q_{i}$ ($q_1=Q_{M}$, and for $i>1$,
$q_{i}=Ze$) located at the center of a hard sphere with radius
$\eta_{i}$ ($\eta_{1}=R_{M}$, and for $i>1$, $\eta_{i}=a$). The
distance between particles $i$ and $j$ is $d_{ij}$. The dielectric
constant is set to $\epsilon=80$ everywhere and there are no
interactions with anything outside of the cell.

Selecting a particle at random, the MC program attempts to
reposition it randomly within a cubic volume of $(3.2$ nm$)^{3}$
centered on the particle's current position. The total electrostatic
energy of the system, $\mathcal {E}$, is calculated after each
attempted move. Modeled as hard spheres, if any of the particles
overlap after an attempted move, such that
$d_{ij}<\eta_{i}+\eta_{j}$ the move is rejected. Additionally, any
attempted move that places a particle outside of the cell,
$r>r_{max}$, is also rejected. Otherwise, moves are accepted or
rejected based on the traditional Metropolis algorithm. Simulations
attempt 52 billion moves, of which $\sim 4\% $ are accepted,
resulting in each particle being moved an average of 20 million
times. This low acceptance rate is due to most of the $Z$-ions being
condensed on the macroion surface where their average separation is
~$b=1.0$ nm; one can increase the rate to $\sim 8\%$ by shrinking
the volume in which the $Z$-ion is randomly repositioned to $(1.6$
nm$)^{3}$. To ensure thermalization, 5 million moves are attempted
before beginning the analysis of $N(r)$, the $Z$-ion's radial
distribution.

Following thermalization, $N(r)$ is computed after every 20,000
attempted moves by dividing the simulation space around the central
macroion into bins that are concentric spherical shells of thickness
$0.1$ nm, counting the ion population within each bin, and then
calculating the average $Z$-ion density of each bin. We now
introduce the empiric mean field potential, $\phi(r)$, which
corresponds to the MC $N(r)$, and is calculated from the radial
distributions of the ions in the following way. First, the electric
field is determined at the outer edge of each spherical shell by
applying Gauss' Law to the integrated charge. Then, the potential
$\phi(r)$ is calculated by discreetly integrating the electric field
in the radial direction. The empiric mean-field potential,
$\phi(r)$, has nothing to do with the PB potential obtained by a
solution of the spherical PB equation because, due to correlation
effects, the MC $N(r)$ differs from Eq. (\ref{eqn:GC}). In the
present case, $Z$-ions are strongly condensed at the surface of the
macroion, and therefore the potential $\phi(r)$ decays so fast with
$r$ that the interaction energy of a stray $Z$-ion, $Ze\phi(r)$,
becomes less than $k_{B}T$ already at $r>5.65$ nm.

The main point of this paper is that the concentration of $Z$-ions,
$N(r)$, at a distance $r$ from the center of the macroion, is only
weakly influenced by the empiric mean-field potential energy
$Ze\phi(r)$ and is mostly determined by the attractive correlation
energy $U_{c}(r)$. We extract $U_{c}(r)-U_c(r_{max})$ from the
simulation data assuming that $Z$-ions that stray from the macroion
surface are Boltzmann distributed according to,
\begin{equation}
N(r) =
N(r_{max})\exp{\left(-\frac{Ze\phi(r)}{k_{B}T}-\frac{U_{c}(r)-U_{c}(r_{max})}{
k_{B}T}\right)}, \label{eqn:BoltzmannZion}
\end{equation}
so that the change in the attractive correlation energy for a
$Z$-ion moved from $r_{max}$ to $r$ is
\begin{equation}
U_{c}(r)-U_{c}(r_{max})=
-k_{B}T\ln\left(\frac{N(r)}{N(r_{max})}\right)-Ze\phi(r),
\label{eqn:CorrPotMC}
\end{equation}
where we took into account that $\phi(r_{max})=0$ because our system
is neutral.

\begin{figure}
\includegraphics[scale=0.50]{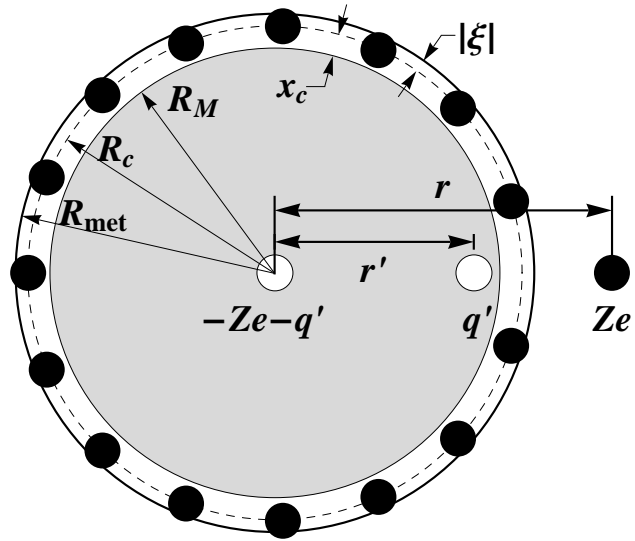}
\caption{The generalization of Fig. \ref{fig:CorrHolePlane} to a
spherical geometry. A stray $Z$-ion with charge $Ze$ is shown in a
cross-sectional view at a distance $r$ from the center of a
spherical macroion with charge $Q_{M}$, which is covered by
condensed $Z$-ions (black spheres). The condensed $Z$-ions are
located at an average distance of $R_{c}\equiv
R_{M}+x_{c}=R_{M}+a+\lambda$ from the center of the macroion. The
stray $Z$-ion makes a correlation hole with charge $-Ze$, where the
concentration of $Z$-ions is depleted. The resulting correlation
potential can be modeled as if the $Z$-ion were near a metallic
sphere with effective radius $R_{met}=R_{c}+|\xi|$. The image
charges, $-Ze$ and $-q'$ located at the center, and $q'$ located at
a distance $r'$ away from the center, are shown by white spheres.}
\label{fig:Macroion_Sphere}
\end{figure}

We need to recalculate the theoretical form of $U_{im}$ for a
spherical macroion geometry (see Fig. \ref{fig:Macroion_Sphere}), to
test that, for $r-R_{met}\gtrsim b$,
\begin{equation}
\Delta U(r)\equiv
\Bigg[U_{c}(r)-U_{c}(r_{max})\Bigg]-\Bigg[U_{im}(r)-U_{im}(r_{max})\Bigg]=0.
\label{eqn:DeltaU}
\end{equation}
It is known~\cite{LL84} that a charge $Ze$ at a distance $r >
R_{met}$ from the center of a conducting sphere with radius
$R_{met}$ and a net charge of $-Ze$, induces two image charges
within the sphere. The charge $q'=-ZeR_{met}/r$ is located at a
distance $r'=R_{met}^{2}/r$ from the sphere's center, and the
compensating charge $-q'$ is located at the center of the sphere.
The net charge of the macroion and the SCL, $-Ze$, accounts for the
departure of the stray $Z$-ion and is also fixed at the center of
the sphere.
In the presence of these three charges a stray $Z$-ion, located at
$r$, has potential energy given by
\begin{eqnarray}
U_{im}(r) = -\frac{(Ze)^{2}}{r \epsilon}+
\frac{Zeq'}{2(r-r')\epsilon}-\frac{Zeq'}{2 r\epsilon}.
\label{eqn:CorrPotTheory}
\end{eqnarray}
The net charge $-Ze$ has fixed magnitude and position because,
unlike charges $q'$ and $-q'$, it is not created by the stray
$Z$-ion polarizing the SCL; therefore, the interaction term that
involves the net charge does not include a factor of $1/2$. In the
limit $x=r-R_{M}\ll R_{M}$, we recover the planar $U_{im}(x)$ of Eq.
(\ref{eqn:PlaneImage}), because $U_{im}(r)$ is dominated by the
influence of charge $q'\simeq-Ze$ located at $r'\simeq R_{met}-r$.

The first term within the parentheses of Eq.
(\ref{eqn:CorrPotTheory}) is written for the case when all but one
of the mobile charges ($Z$-ions) are located, as in metal, at the
surface. This term then describes a stray $Z$-ion's attraction to
the fraction of $Q_M$ left uncompensated due to its departure. In
other words, this term is used to exclude the stray $Z$-ion's self
interaction with its contribution to the mean-field
potential~\cite{CenterCharge}.

To compare Eq. (\ref{eqn:CorrPotTheory}) to the simulation in the
next section, we take $R_{met}=R_{M}+x_{c}\equiv R_{c}$, which
aligns the metallic surface with the average position of the centers
of the $Z$-ions that comprise the SCL (see Fig.
\ref{fig:Macroion_Sphere}). Because our macroion is a sphere and not
a plane, the magnitude of its electric field drops as $E \propto
1/r^{2}$ at $r>R_{M}$. Therefore, $E=2 \pi \sigma/\epsilon$, used to
calculate $\lambda$, should be modified slightly since the $Z$-ion's
centers are never closer than $a$ to the macroion's surface. We
introduce $\sigma_{s}=\sigma[R_{M}/(R_{M}+a)]^{2}$ to correct the
electric field. This leads to, $\sigma_{s}=0.819 $ $e/$nm$^{2}$,
$\lambda_{s}=0.0913$ nm, $\Gamma_{s}=5.9$, and $R_{met}=5.49$ nm.

\section{Results of MC simulation}\label{sec:MCResults}

\begin{figure}[h!]
\includegraphics[scale=0.472]{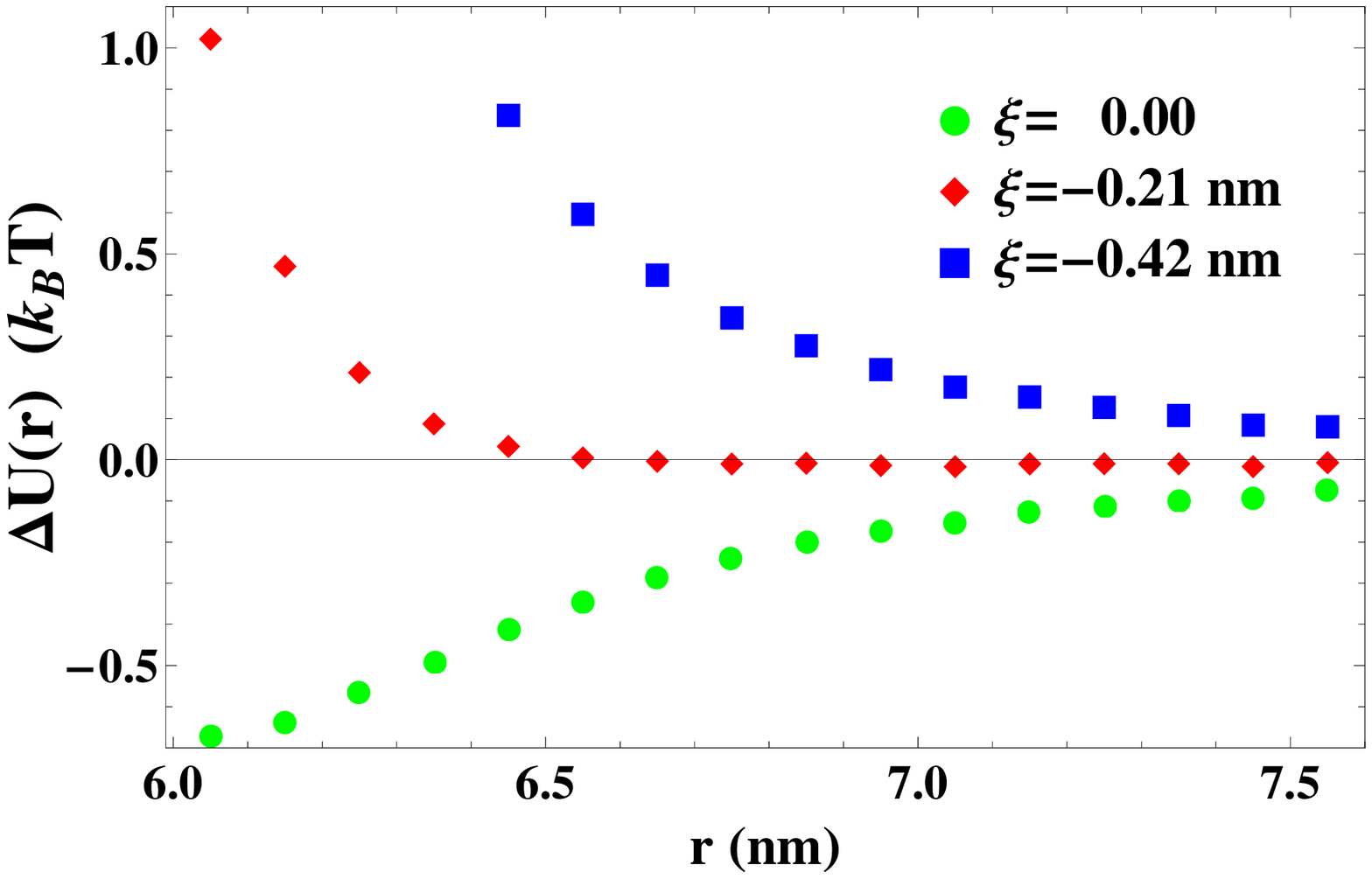}
\caption{The difference, $\Delta U(r)$ (Eq. (\ref{eqn:DeltaU})),
between the correlation attraction energy extracted from the MC
simulation and the result of the image theory, as a function of a
stray $Z$-ion's distance from the center of the macroion for three
different values of the adsorbed $Z$-ion's screening length, $2\xi$.
The length, $|\xi|$, determines the increased radius,
$R_{met}=R_{M}+a+\lambda+|\xi|$, of the effective metallic sphere
used to calculate $U_{im}(r)$ (Eq (\ref{eqn:CorrPotTheory})). The
green circles correspond to $\xi=0$, assumed in the original theory
of Refs.~\cite{PS99,S99}. The red diamonds correspond to the best
fit to zero, $\xi=-0.21$ nm. The blue squares correspond to,
$\xi=-0.42$ nm and are shown for comparison.} \label{fig:MCdiff}
\end{figure}

$\Delta U(r)$ (Eq. (\ref{eqn:DeltaU})), the difference between the
attractive correlation energy extracted from the MC simulation and
the result of the image theory is plotted in Fig. \ref{fig:MCdiff}
for $R_{met}=5.49$ nm (green circles). As expected, when
$r-R_{met}\lesssim b$, i.e. at $r\lesssim 6.5$ nm the difference is
significantly less than zero since in this range the SCL does not
function well as a metal due to its discreteness, and, therefore,
the attractive correlation energy, $U_{c}(r)$, saturates. However,
there is also weaker disagreement for $r \gtrsim 6.5$nm, which
decreases with distance from the macroion. This suggests that we
have improperly identified the radius of the effective metallic
sphere used to calculate $U_{im}(r)$. To allow for the adjustment of
$R_{met}$, we introduce the length $|\xi|$ so that
\begin{equation}
R_{met}=R_{c}+|\xi|. \label{eqn:Rmet}
\end{equation}
By minimizing the root-mean-square of $\Delta U(r)$ with respect to
$|\xi|$, on the interval $6.4$ nm $\leq r \leq 7.4$ nm, we
determined that $|\xi|\simeq 0.21$ nm provides the best fit for
$\Delta U(r)=0$. The quality of this fit is illustrated in Fig.
\ref{fig:MCdiff} (red diamonds). This small correction $|\xi|$ to
$R_{c}$ indicates that the foundation of Eq.
(\ref{eqn:CorrPotTheory}), the attractive image interaction, is
sound.
\begin{figure}[h!]
\includegraphics[scale=0.508]{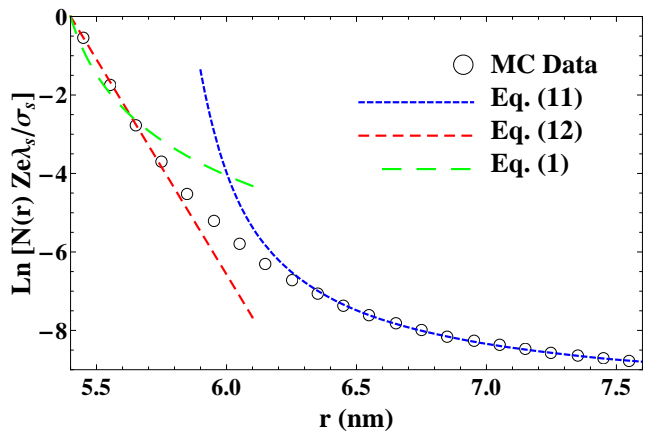}
\caption{Concentration of $Z$-ions, $N(r)$, as a function of
distance from the center of the macroion, starting at $R_M+a=5.4$
nm. The circles represent the data from the MC simulation. The
result of the image theory, Eq. (\ref{eqn:BoltzmannZionImage}), is
shown by short, blue dashes. The medium length, red dashes show Eq.
(\ref{eqn:closeSphere}). The long, green dashes show the
Gouy-Chapman solution (Eq. (\ref{eqn:GC})), with $\lambda_{s}$
substituted for $\lambda$. The error bars for the MC data are
smaller than the size of the symbols.}
\label{fig:MCvsTheory_CorrPot_Plot}
\end{figure}

In Sec. \ref{sec:Location}, we analytically calculate $U_{im}(x)$ in
order to find the necessary adjustment in $R_{met}$ by considering
the response of a SCL, made up of adsorbed $Z$-ions on a planar
macroion, to the presence of a single stray $Z$-ion above the SCL.
It is determined that the SCL screens the potential of the stray
$Z$-ion with a negative screening length, $2\xi$, where $\xi=-0.20$
nm. This moves the effective metallic surface further away from the
macroion's surface by $|\xi|=0.20$ nm in reasonable agreement with
the MC data (see Fig. \ref{fig:CorrHolePlane} and Fig.
\ref{fig:Macroion_Sphere}).

In Fig. \ref{fig:MCvsTheory_CorrPot_Plot}, the concentration $N(r)$
obtained from the MC simulation is compared to
\begin{equation}
N(r) =
N(r_{max})\exp{\left(-\frac{Ze\phi(r)}{k_{B}T}-\frac{U_{im}(r)-U_{im}(r_{max})}{k_{B}T}\right)},
\label{eqn:BoltzmannZionImage}
\end{equation}
which uses $\xi=-0.21$ nm to calculate $U_{im}(r)$ (both $\phi(r)$
and $N(r_{max})$ are obtained from the MC simulation). The agreement
between the MC data and Eq. (\ref{eqn:BoltzmannZionImage}) is
obvious when $r \gtrsim 6.5$ nm. In Fig.
\ref{fig:MCvsTheory_CorrPot_Plot}, we also compare Eq.
(\ref{eqn:close}), modified for a spherical geometry,
\begin{equation}
N(r)=\frac{\sigma_{s}}{Ze\lambda_{s}} \exp\left[-\frac{(r-R_{M}-
a)}{\lambda_{s}}\right], \label{eqn:closeSphere}
\end{equation}
to the MC concentration data. At small distances, $r-R_{M}+a\lesssim
b$, i.e. $r\lesssim 5.8$ nm, we find good agreement with the
exponential decay predicted in Refs.~\cite{PS99,S99} and confirmed
in Refs.~\cite{MN2000, BAO2004, MN2001}.

Let us now comment on what happens at larger distances from the
macroion, which are not shown in Fig.
\ref{fig:MCvsTheory_CorrPot_Plot} and can not be studied well with
the small size of the simulation cell used in this paper. According
to Refs.~\cite{PS99,S99}, at distances larger than
\begin{equation}
\Lambda = \left(\frac{ e \lambda}{2\pi Z \sigma l_B
}\right)^{1/2}\exp\left(\frac{|\mu|}{2k_{B}T}\right)
\label{eqn:NewGClength}
\end{equation}
from the planar macroion the PB approximation takes over and
\begin{equation}
N(x)=\frac{1}{2 \pi Z^{2} l_{B} }\frac{1}{(\Lambda + x - a)^{2}}.
\label{eqn:NewGC}
\end{equation}
Here $\mu$ is the chemical potential of a Z-ion in a SCL. It has
been shown that for a SCL on a charged background (one-component
plasma), at $1<\Gamma<15$, $\mu$ is approximated well
by~\cite{HT78,S99},
\begin{equation}
\mu=-k_{B} T(1.65\Gamma-2.61\Gamma^{1/4}+0.26\ln{\Gamma} + 1.95),
\label{eqn:chempot}
\end{equation}
where the first term of this expansion corresponds to the chemical
potential of a Wigner Crystal. For our parameters, $Z=3$ and
$\sigma=\sigma_{s}=0.819 e$ nm$^{-2}$, this leads to the length
$\Lambda=5.18$nm. Then, the approximate extension of Eq.
(\ref{eqn:NewGC}) to the spherical geometry using $x = r- R_M$ at $r
= 7.55$ nm gives $\ln[N(r)Ze\lambda_s/\sigma_s]= -8.65$, very close
to the MC result $-8.77$ (see Fig.
\ref{fig:MCvsTheory_CorrPot_Plot}). The idea behind the results of
Eq. (\ref{eqn:NewGClength}) and Eq. (\ref{eqn:NewGC}) is that the
correlation physics at small distances $x-x_c \ll \Lambda$, produces
a new boundary condition on the concentration of $Z$-ions for the
long distance solution of the PB equation~\cite{PS99,S99}.

The authors of a recent paper~\cite{Levin} have already studied
$N(r)$ at large distances by MC simulation in a much a larger
spherical cell and found that it is in agreement with the
predictions of the PB approach based on the correlation driven
boundary condition. They, however, did not identify the image
domain of distances $r$ which we concentrate here upon. Thus, all
three asymptotic regimes, predicted in Refs.~\cite{PS99,S99},
namely Eq. (\ref{eqn:close}) at $x-a < b$, Eq.
(\ref{eqn:PlaneImage}) at  $b < x-a \ll \Lambda$, and Eq.
(\ref{eqn:NewGC}) at $x-a > \Lambda$ are now confirmed by MC
simulations.

\begin{figure}[t!]
\includegraphics[scale=0.505]{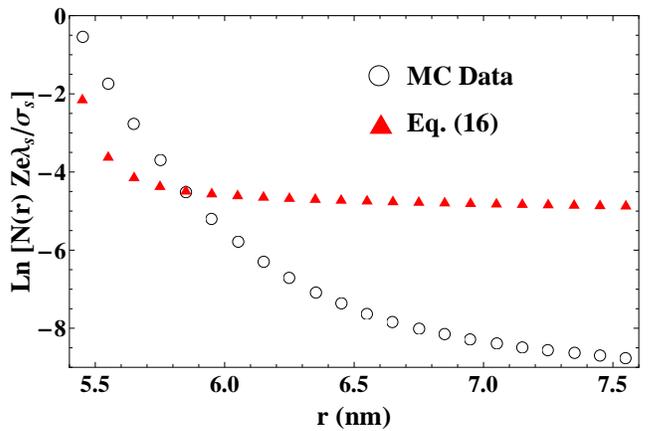}
\caption{Concentration of $Z$-ions, $N(r)$, as a function of
distance from the center of the macroion. The MC concentration
(circles; the same data as in Fig.
\ref{fig:MCvsTheory_CorrPot_Plot}) differs strongly from the
concentration of $Z$-ions (red triangles) obtained from Eq.
(\ref{eqn:meanfield}) with the empiric mean-field potential
potential $\phi(r)$. } \label{fig:MCvsMean}
\end{figure}

We have shown above that the standard mean-field
theory~\cite{Gouy1910,Chap1913} fails to describe screening by
multivalent counterions. Let us now show that another mean-field
approximation, which we call the empiric mean-field, fails more
dramatically. The empiric mean-field potential, $\phi(r)$, was
introduced in Sec. \ref{sec:MCSimulation} and is obtained using the
distribution of charge realized in our MC simulation. In Fig.
\ref{fig:MCvsMean}, we compare the $Z$-ion concentration obtained
from the MC simulation to the $Z$-ion concentration predicted using
the empiric mean-field potential,
\begin{equation}
N(r)=N_{0}\exp \left(-\frac{Ze\phi(r)}{k_{B}T}\right).
\label{eqn:meanfield}
\end{equation}
Here, $N_{0}=2.18 \times 10^{-2}$ nm$^{-3}$ is the concentration
necessary to normalize the number of $Z$-ions, in the range $5.4$ nm
$<r<10.0$ nm, to 100. Clearly, the empiric mean-field potential is
not self-consistent; Eq. (\ref{eqn:meanfield}) predicts that there
are many more $Z$-ions, at $r>6.0$ nm, than are actually present in
the distribution that produced $\phi(r)$. The distribution of the
$Z$-ions, for $r-R_{met}\gtrsim b$, is strongly influenced by the
attractive correlation interaction and, therefore, cannot be
predicted by the empiric mean-field interaction alone.

\section{Theory of image potential and effective metallic surface}\label{sec:Location}
In this section we return to the plane geometry of Fig.
\ref{fig:CorrHolePlane} and analytically derive Eq.
(\ref{eqn:PlaneImage}) for $U_{im}(x)$. In the process of this
derivation, we find the theoretical location, $x_{met}$, of the
effective metallic surface. The probe charge, a stray $Z$-ion, is
positioned far above the plane at $x'=x \gg b$ and $\varrho = 0$,
where $x'$ is the axis and $\varrho$ is the radius of the
cylindrical coordinate system ($x', \varrho,\theta$). A SCL of
$Z$-ions is located in the ($\varrho$,$\theta$) plane at $x'=x_{c}$,
where the typical distance that separates adjacent $Z$-ions is $b$.

The plan of this section consists of, (1) determining the analytic
solution for the total potential of the system, $\psi(\varrho,x')$,
(2) presenting it as a sum of two potentials: one of the stray
$Z$-ion and the other of the induced charge density of the SCL,
$\psi_{ind}(\varrho,x')$ (the potential of a point like image) and
(3) finding the position of the effective metallic plane, $x_{met}$,
so that the attractive interaction energy
$\frac{1}{2}Ze\psi_{ind}(0, x) = U_{im}(x)$. Below, we show that
$x_{met}=x_{c} - \xi$, where $2\xi$ is the screening length of the
SCL, which we also calculate.

To find the potential $\psi(\varrho,x')$ we solve Poisson's
equation,
\begin{equation}
\nabla^{2}\psi\left(\varrho,x'\right)=-\frac{4
\pi}{\epsilon}\rho(\varrho,x'), \label{eqn:PoissonsEqn}
\end{equation}
where
$\rho(\varrho,x')=\rho_{ext}(\varrho,x')+\rho_{ind}(\varrho,x')$,
with $\rho_{ext}= Z e \delta(\varrho)\delta(x'-x)/(\pi \varrho)$,
and the charge density that is induced within the SCL is given by
\begin{eqnarray}
\rho_{ind}(\varrho,x')&=&Ze\left[n\Bigg(\psi(\varrho,x_{c})\Bigg)-n(0)\right]\delta(x'-x_{c})
\nonumber \\
&=& Ze\psi(\varrho,x_{c})\frac{d n}{d\psi}\delta(x'-x_{c})
\nonumber \\
&=&-(Ze)^{2}\psi(\varrho,x_{c})\frac{d n}{d \mu}\delta(x'-x_{c}).
\label{eqn:induce charge density linearized}
\end{eqnarray}
Here, $n(\psi)$, is the number of $Z$-ions per unit area as a
function of the local total potential, $\psi(\varrho,x_{c})$, and
$\mu$ is the chemical potential of the SCL. We consider the case,
$x-x_{c}\gg b$, when the stray Z-ion produces a weak potential in
the $x'=x_{c}$ plane ($Ze\psi(\varrho,x_{c})/k_{B}T\ll1$). This
allows us to linearize $\varrho_{ind}$ with respect to $\psi$, in
Eq. ({\ref{eqn:induce charge density linearized}). Rewriting Eq.
(\ref{eqn:PoissonsEqn}) with the help of Eq. (\ref{eqn:induce charge
density linearized}) results in
\begin{equation}
\nabla^{2}\psi\left(\varrho,x'\right)=-\frac{4
\pi}{\epsilon}\rho_{ext}(\varrho,x')
+\frac{1}{\xi}\psi(\varrho,x_{c})\delta(x'-x_{c}),
\label{eqn:PoissonsEqn rewritten}
\end{equation}
where,
\begin{equation}
\xi= \frac{\epsilon}{4 \pi (Ze)^{2}}\frac{d\mu}{dn}=\frac{1}{2
\kappa}, \label{eqn:q_2}
\end{equation}
and $\kappa$ is the inverse screening length of the SCL of adsorbed
$Z$-ions~\cite{Efros08, AFS82}.

In order to calculate $\xi$ we use $\mu(n)$ as given by Eq.
(\ref{eqn:chempot}) and the definition of $\Gamma$ from Eq.
(\ref{eq:Gamma}) and $b=(\pi n)^{-1/2}$.  For $Z=3$, $\sigma =
\sigma_{s} =0.819$ $e/$nm$^{2}$ and $\epsilon = 80$, we find that
$\xi = -0.20$ nm.

In order to solve Eq. (\ref{eqn:PoissonsEqn rewritten}) for
$\psi(\varrho,x')$, we substitute
\begin{equation}
\psi(\varrho,x')=\int^{\infty}_{0} k A_{k}(x')J_{0}(k\varrho)dk,
\label{eqn:Fourier Bessel Expantion}
\end{equation}
into Eq. (\ref{eqn:PoissonsEqn rewritten}), where $A_{k}(x')$ are
the coefficients of the expansion and $J_{0}(k\varrho)$ is the
zeroth order Bessel function. This yields~\cite{AFS82}
\begin{eqnarray}
&{}& \psi(\varrho,x') =
\frac{Ze}{\epsilon}\frac{1}{\sqrt{\left(x-x'\right)^{2}+\varrho^{2}}}-
\nonumber \\
& & \frac{Ze}{\epsilon}\int^{\infty}_{0} \frac{1}{2k \xi+1}
\exp{[-k(x'+x-2 x_{c})]}J_{0}(k\varrho)dk, \nonumber
\\
\label{eqn:Pot Solution}
\end{eqnarray}
Because the screening length $\xi<0$, the second term diverges. To
obtain a solution despite this pole, following Ref.~\cite{Efros08},
we consider the contribution to $\psi(\varrho, x')$ from $k\ll
1/|\xi|$, only. Such an approach is valid if the stray $Z$-ion, and
the observation point $x'$, are a large distance away from the SCL:
$(x-x_{c}), (x'-x_{c})\gg |\xi|$. This allows us to expand
$1/(2k\xi+1)$ in Eq. (\ref{eqn:Pot Solution}) around $k=0$, so that
$1/(2k\xi+1)\simeq 1-2k\xi$, and we arrive at
\begin{eqnarray}
&{}&\psi(\varrho,x')=\frac{Ze}{\epsilon\sqrt{\varrho^{2}+\left(x-x'\right)^{2}}}
\nonumber \\
&&-\frac{Ze}{\epsilon\sqrt{\varrho^{2}+(x'+x-2x_{c})^{2}}}+\frac{2(x'+x-2x_{c})Ze\xi}{\epsilon
\left[\varrho^{2}+(x'+x-2x_{c})^{2}\right]^{3/2}}. \nonumber
\\
\label{eqn:Pot Approx Solution}
\end{eqnarray}
The first term of Eq. (\ref{eqn:Pot Approx Solution}) is the
potential created directly by the stray $Z$-ion. The other two terms
are the first two terms of the expansion of the induced potential,
$\psi_{ind}(\varrho,x')$, with respect to $\xi$. We are now in a
position to recast $\psi_{ind}(0, x)$ at $(x-x_{c})\gg |\xi|$, as
being created by an image charge a distance $s$ below the stray
$Z$-ion,
\begin{eqnarray}
\frac{1}{2}Ze \psi_{ind}(0,x) &=&
-\frac{(Ze)^{2}}{4(x-x_{c})\epsilon}+\xi \frac{(Ze)^{2}}{ 4
(x-x_{c})^{2}\epsilon} \nonumber \\
& \simeq & -\frac{(Ze)^{2}}{2s \epsilon}=U_{im}(x),
\label{eqn:Metallic and induced pot at z0}
\end{eqnarray}
where, $s=2(x-x_{c}+\xi)$. Specifying that the metallic plane must
lie half way between the real charge and the image charge sets its
position at $x_{met}=x-s/2=x_{c}-\xi$. Therefore the effective
metallic plane is found a distance $\xi$ above the plane of the
adsorbed $Z$-ion's centers (Fig. \ref{fig:CorrHolePlane}). This
agrees with the statement of Ref.~\cite{Efros08}, that the potential
created by the stray $Z$-ion is negative in the $x'=x_{c}$ plane.
The theoretical value $\xi=-0.20$ nm is in reasonable agreement with
our MC result, $\xi=-0.21$ nm (Sec \ref{sec:MCResults}). Moreover,
we have demonstrated that the image attraction predicted in
Ref.~\cite{PS99, S99} can be derived analytically in the limit $x
\gg b$.

\section{Screening the image by adding 1:1 salt}\label{sec:addedsalt}
In this section we modify our system to include a small
concentration of 1:1 salt molecules such as NaCl. By taking into
account the effect of screening on the attractive interaction energy
between a stray $Z$-ion and its image, $U_{im}$, we obtain a new
prediction for the concentration of $Z$-ions, $N(r)$, which is in
reasonable agreement with the new MC results. In calculating
$U_{im}$ we assume that the concentration $c$ of 1:1 salt is so
small that the total potential $\psi$ at any point in the bulk
solution ($r \gtrsim 6.0$ nm) obeys the linearized Poisson-Boltzmann
equation,
\begin{equation}
\nabla^2 \psi=\kappa_{b}^{2} \psi,
\label{eqn:LinearizedPBeqn}
\end{equation}
where $1/\kappa_{b}$ is the Debye-H\"{u}ckel (DH) screening length,
\begin{equation}
\frac{1}{\kappa_{b}}=\sqrt{\frac{\epsilon k_{B} T}{8 \pi  e^{2} c}}.
\label{eqn:DHscreenLength}
\end{equation}
The exact solution of Eq. (\ref{eqn:LinearizedPBeqn}) for a point
charge a distance $r-R_{met}$ away from the surface of a grounded
metallic sphere in a weak electrolyte is known~\cite{HO1994};
however, we will avoid the complexity of this solution and
approximate the spherical macroion and its SCL as a grounded
metallic plane. As seen in Fig.
\ref{fig:MCvsTheory_CorrPotScreened_Plot}, where Eq.
(\ref{eqn:PlaneImage}) is used to calculate $N(r)$ (short blue
dashes) for $c=0$, one obtains reasonable agreement with the MC
$N(r)$ (circles) without using Eq. (\ref{eqn:CorrPotTheory}) as we
did in Sec. \ref{sec:MCResults}. This demonstrates that the
influence of the total central charge, $-Ze-q'$, is very small. The
reason for this is that when a stray $Z$-ion is close to the
macroion surface, the total central charge is much smaller than the
image charge $q'$. Additionally, the central charge is much farther
from the stray $Z$-ion than the image charge $q'$. When the system
includes 1:1 salt, the influence of the total central charge is
further reduced due to screening.
\begin{figure}[h!]
\includegraphics[scale=0.51]{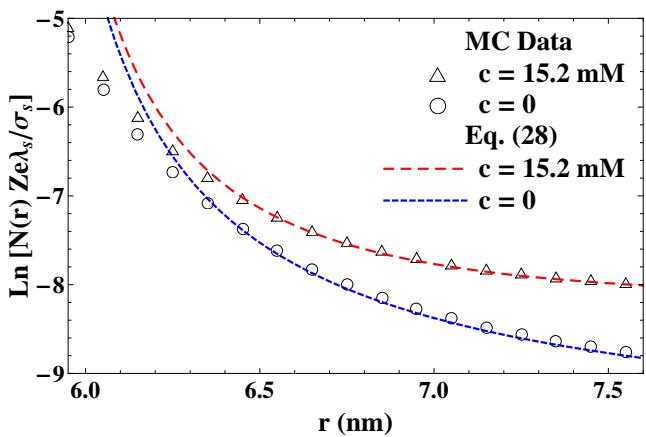}
\caption{Concentration of $Z$-ions, $N(r)$, in a weak electrolyte
solution, as a function of distance from the center of the macroion.
The shapes represent the data from the MC simulations for two
different concentrations of 1:1 salt: 15.2 mM (triangles), and 0
(circles). The result of the screened image theory, Eq.
(\ref{eqn:BoltzmannZionImageScreened}) with $x=r-R_{met}+x_{met}$,
is shown by medium length red dashes for c=15.2 mM, and short blue
dashes for c=0. The error bars for the MC data are smaller than the
size of the symbols.} \label{fig:MCvsTheory_CorrPotScreened_Plot}
\end{figure}

Consider a $Z$-ion which is submerged in a weak electrolyte solution
with dielectric constant $\epsilon$, and is a distance $x-x_{met}$
above a grounded metallic plane located at $x_{met}$ (see Fig.
\ref{fig:CorrHolePlane}). For this system, the solution to Eq.
(\ref{eqn:LinearizedPBeqn}), subject to the boundary condition
$\psi(x_{met})=0$, can be found using the method of
images~\cite{HO1994}. To satisfy the boundary condition, the image
potential must exactly cancel the potential of our $Z$-ion in the
metallic plane. Such an image potential is provided by the DH
screened potential of a charge ($-Ze$) located at $x'=-x+2x_{met}$.

The interaction energy of a stray $Z$-ion with its screened image is
now readily calculated and is given by,
\begin{equation}
U_{im}(x)=-\frac{Z^{2} e^{2}}{4 \epsilon
(x-x_{met})}\exp(-2\kappa_{b} [x-x_{met}]).
\label{eqn:PlaneImageScreened}
\end{equation}
In the limit of infinite dilution $c \rightarrow 0$ , or
equivalently $\kappa_{b} \rightarrow 0$, Eq.
(\ref{eqn:PlaneImageScreened}) is equal to Eq.
(\ref{eqn:PlaneImage}), as expected. Onsager and
Samaras~\cite{OS1934} obtained the same result, but with the
opposite sign, for an ion's interaction with its image at an
electrolyte-air interface resulting in a repulsive force. To compare
Eq. (\ref{eqn:PlaneImageScreened}) to the spherical geometry of our
MC simulation, we take $x=r+x_{met}-R_{met}$ using
$R_{met}=R_{c}+|\xi|$.

The MC simulation described in Sec. \ref{sec:MCSimulation} was
modified to include $M$ 1:1 salt molecules, resulting in $M$ ions of
charge $e$ and $M$ ions of charge $-e$. All of the monovalent salt
ions have their charge at the center of a hard sphere with radius
$\eta=0.2$ nm. We studied a 1:1 salt concentrations of $15.2$ mM
corresponding to the addition of $M=34$ salt molecules to the
solution. The following changes were made to the MC simulation to
properly incorporate the new ions. The sum used to calculate the
total electrostatic energy of the system (Eq. (\ref{eqn:coulomb}))
was changed to include the monovalent ions, and the monovalent ions
were also incorporated into the calculation of the empiric
mean-field potential $\phi(r)$.

In Fig. \ref{fig:MCvsTheory_CorrPotScreened_Plot}, the concentration
N(r) obtained from the MC simulations is compared to
\begin{eqnarray}
N(r) =
N(r_{0})\exp  \bigg( - \frac{Ze[\phi(r)-\phi(r_0)]}{k_{B}T} \nonumber \\
  - \frac{U_{im}(r)-U_{im}(r_{0})}{k_{B}T} \bigg),
\label{eqn:BoltzmannZionImageScreened}
\end{eqnarray}
which uses $\xi=-0.21$ nm to calculate $U_{im}(x)$ (Eq.
(\ref{eqn:PlaneImageScreened})) (both $\phi(r)$ and $N(r_{0})$ are
obtained from the MC simulations). Because there are no screening
particles outside of the simulation cell, $Z$-ions near the wall of
the cell are repelled from this interface~\cite{N1999} even though
there is not a jump in the dielectric constant. To keep this effect
separate from the image interaction of interest, we chose our
reference point at $r_{0}=8.05$ nm instead of $r_{max}$. Because the
stray $Z$-ion's attraction to the image is reduced by screening, we
see that in Fig. \ref{fig:MCvsTheory_CorrPotScreened_Plot} the
concentration of $Z$-ions is higher at distances $r
> 60$ nm when 1:1 salt is present in the solution. Even with
the addition of 1:1 salt to the solution the agreement between the
MC data and Eq. (\ref{eqn:BoltzmannZionImageScreened}) for $r
\gtrsim 6.5$ nm is reasonable, demonstrating that the metallic
behavior of the SCL on the macroion surface survives, and that the
image attraction is still important to determining the $Z$-ion's
concentration.

\section{Conclusion}

To summarize, we have studied the role of correlations among
adsorbed $Z$-ions in attracting stray $Z$-ions and influencing their
distribution in the screening atmosphere. Adsorbed $Z$-ions on the
surface of the macroion form a strongly correlated liquid (SCL). The
SCL acts as an effective metallic surface for $Z$-ions that stray
from the macroion surface to distances larger than the average
distance between $Z$-ions of the SCL. Using Monte Carlo (MC)
simulations, we verified the theoretical prediction~\cite{PS99, S99}
that a stray $Z$-ion is attracted to its electrostatic image created
behind the effective metallic surface. As a small correction to
Refs.~\cite{PS99, S99}, however, our MC simulation showed that the
effective metallic surface is not aligned with the average position
of the adsorbed $Z$-ion's centers, but is slightly above the
adsorbed $Z$-ion's centers. This offset was calculated analytically
to be $|\xi|$, where $2\xi$ is the screening length of the SCL. Our
analytic theory is in reasonable agreement with the MC data.
Extending the original image theory of Refs.~\cite{PS99, S99}, we
demonstrated that the attractive image interaction, while screened,
persists in a weak electrolyte.

In Ref.~\cite{Colloquium02} the attractive image interaction, which
we have studied here, was used to interpret the origin of the
negative chemical potential of the condensed $Z$-ions (Eq.
\ref{eqn:chempot}). As a stray $Z$-ion, attracted to its image,
approaches the surface of the macroion it reaches a distance, $r
\sim R_{M}+b$, where the SCL fails to act as a good metal and the
correlation attraction $U_{c}(r)$ saturates at $\mu \sim -Z^2 e^2/ 4
\epsilon b$; this saturation can be seen in Fig.
\ref{fig:MCvsTheory_CorrPot_Plot} as the growth of $\Delta U(r)$ at
$r<6.5$ nm. It is this negative chemical potential, brought about by
the attractive image interaction, at $r\gtrsim R_{met} + b$, which
drives charge inversion (over compensation of the macroion's bare
charge with condensed $Z$-ions), a phenomenon that has generated
much interest~\cite{PS99, S99, Colloquium02, L2002, N_rev2005,
Mess_rev2008, MN2000, BAO2004, MN2001}. Thus, we believe that this
paper helps to clarify the origin of charge inversion.

\section*[\nonumber]{Acknowledgments}

We are grateful to T. T. Nguyen for his help in writing the MC code,
to A. L. Efros and Y. Levin for sharing their
preprints~\cite{Efros08,Levin} and useful comments on the
manuscript, to A. Yu. Grosberg for useful discussions, and to B.
Skinner for proofreading the draft. M.S.L. wishes to thank FTPI of
the University of Minnesota for financial support.


\begin{thebibliography}{99}

\bibitem[*]{contact} loth@physics.umn.edu

\bibitem{Gouy1910} M. Gouy, J. Phys. Théor. Appl. {\bf 9}, 457 (1910).

\bibitem{Chap1913} D. L. Chapman, Philos. Mag. {\bf 25}, 475 (1913).

\bibitem{PS99} V. I. Perel and B. I. Shklovskii, Physica {\bf A 274}, 466 (1999).

\bibitem{S99} B. I. Shklovskii, Phys. Rev. E \textbf{60}, 5802 (1999).

\bibitem{Colloquium02} A. Yu. Grosberg, T. T. Nguyen, and B. I. Shklovskii, Rev. Mod. Phys. \textbf{74}, 329 (2002).

\bibitem{L2002} Y. Levin, Rep. Prog. Phys. {\bf 65}, 1577 (2002).

\bibitem{N_rev2005} H. Boroudjerdi, Y. W. Kim, A. Naji, R. R. Netz, X. Schlagberger, and A. Serr, Physics Reports {\bf 416}, 129 (2005).

\bibitem{Mess_rev2008} R. Messina, J. Phys.: Condens. Matter 21, 113102 (2009)

\bibitem{MN2000} A. G. Moreira and R. R. Netz, Europhys. Lett. {\bf 52}, 705 (2000)

\bibitem{BAO2004} Y. Burak, D. Andelman, H. Orland Phys. Rev. E {\bf 70}, 016102 (2004)

\bibitem{MN2001} A. G. Moreira and R. R. Netz, Phys. Rev. Lett. {\bf 87}, 078301 (2001).

\bibitem{ROUZ1996} I. Rouzina and V. A. Bloomfield, J. Phys. Chem. {\bf 100}, 9977 (1996).

\bibitem{GRON97} N. Gronbech-Jensen, R. J. Mashl, R. F. Bruinsma, and W. M. Gelbart, Phys. Rev. Lett. {\bf 78}, 2477 (1997).

\bibitem{S1999_2} B. I. Shklovskii, Phys. Rev. Lett. \textbf{82}, 3268 (1999).

\bibitem{BLES1981} M.S. Bello, E.I. Levin, B.I. Shklovskii, A.L. Efros, Sov.  Phys.-JETP {\bf  53}, 822 (1981)

\bibitem{KRAV1990} S. V. Kravchenko, D. A. Rinberg, S. G. Semenchinsky, and V. Pudalov, Phys. Rev. B {\bf 42}, 3741 (1990).

\bibitem{Eis1992} J. P. Eisenstein, L. N. Pfeifer, and K. W. West, Phys. Rev. Lett. {\bf 68}, 674 (1992).

\bibitem{Efros08} A. L. Efros, Phys. Rev. B, \textbf{78}, 155130 (2008).

\bibitem{holm07} O. Lenz, and C. Holm, Euro. Phys. J.E. {\bf 26}, 191 (2008).

\bibitem{Diehl06} A. Diehl, Y. Levin, J. Chem. Phys. {\bf 125}, 054902 (2006).

\bibitem{Diehl08} A. Diehl and Y. Levin, J. Chem. Phys. {\bf 129} 124506 (2008).

\bibitem{Perez05} M. Quesada-Perez, A. Martin-Molina, R. Hidalgo-Alvarez, Langmuir, {\bf 21}, 9231 (2005)

\bibitem{Perez08} A. Martin-Molina, J. A. Maroto-Centeno, R. Hidalgo-Alvarez, M. Quesada-Perez,  J. Chem. Phys. {\bf 125} 124506 (2008).

\bibitem{LL84} L. D. Landau and E. M. Lifshitz, \textit{Electrodynamics of Continuous Media} (Butterworth-Heinenann, 1984), chapter I.

\bibitem{CenterCharge} Actually, $N(r)$ has a tail at $r> R_M+a$. As a result, when a
stray $Z$-ion is located at $r> R_M+a$, depletion of the mean
distribution not only occurs at the surface of the macroion, but a
small fraction, $\delta$, of the total depletion also occurs at
distances larger than $r$. For $r = 6.1$ nm this fraction is $0.02$.
As a result, the absolute value of this interaction energy is
smaller than $Z^{2}e^{2}/r$ by $\sim 2\%$. In Eq.
(\ref{eqn:CorrPotTheory}) and below we neglect this small effect.

\bibitem{Levin} A.P. dos Santos, A. Diehl, and Y. Levin, J. Chem. Phys.

\bibitem{HT78} H. Totsuji, Phys. Rev. A, \textbf{17}, 399 (1978).

\bibitem{AFS82} T. Ando, A. B. Fowler, and F. Stern, Rev. Mod. Phys. \textbf{54}, 437 (1982).

\bibitem{HO1994} H. Ohshima, Adv. Colloid Interface Sci. {\bf 53}, 77 (1994).

\bibitem{OS1934} L. Onsager and N. N. T. Samaras, J. Chem. Phys. {\bf 2}, 528 (1934).

\bibitem{N1999} R. R. Netz, Phys. Rev. E, {\bf 60}, 3174 (1999).


\end{thebibliography}
\end{document}